\documentstyle[12pt]{article}

\begin{document}

\begin{center}
{\large BERRY PHASE EFFECTS IN THE DYNAMICS OF DIRAC ELECTRONS \\[0pt]
IN DOUBLY SPECIAL RELATIVITY FRAMEWORK }\\[0pt]

\vskip 1.5cm Pierre Gosselin \\[0pt]
\vskip 0.5cm Institut Fourier, UMR 5582 CNRS-UJF, UFR de Math\'ematiques, \\[%
0pt]
Universit\'e Grenoble I, \\[0pt]
BP74, 38402 Saint Martin d'H\`eres, Cedex,\\[0pt]
France.

\vskip 1.5cm Alain B\'erard, and Herv\'e Mohrbach\\[0pt]
\vskip 0.5cm Laboratoire de Physique Mol\'eculaire et des Collisions,
ICPMB-FR CNRS 2843,\\[0pt]
Universit\'e Paul Verlaine-Metz,\\[0pt]
57078 Metz Cedex 3,\\[0pt]
France.

\vskip 1.5cm Subir Ghosh\\[0pt]
\vskip 0.5cm Physics and Applied Mathematics Unit,\\[0pt]
Indian Statistical Institute,\\[0pt]
203 B. T. Road, Calcutta 700108, \\[0pt]
India.
\end{center}

\vskip 1cm {\bf Abstract:}\newline

We consider the Doubly Special Relativity (DSR) generalization of Dirac
equation in an external potential in the Magueijo-Smolin base. The particles
obey a modified energy-momentum dispersion relation. The semiclassical
diagonalization of the Dirac Hamiltonian reveals the intrinsic Berry phase
effects in the particle dynamics.

\vskip 2cm \noindent Keywords: $\kappa$-Minkowski space-time, DSR, Foldy
Wouthuysen

\vskip 1cm \noindent PACS Numbers: \newpage

Evidence (see \cite{am1} for discussion and references) of ultra-high energy
cosmic ray particles that violate the Greisen-Zatsepin-Kuzmin bound have
compelled theorists to generalize the conventional energy-momentum
dispersion law of particles,
\begin{equation}
p^2=m^2,  \label{sr}
\end{equation}
based on principles of Special theory of Relativity (SR).
{\footnote{However, it should be pointed out that very recent data
from the Auger cosmic ray observatory  does not quite
support the observation of ultra-high energy cosmic ray particles
although results on the ultra-high energy cosmic photons is still
awaited.}} The extension requires another observer independent
dimensional parameter, apart from $c$, the velocity of light. The
second parameter $(\kappa )$ is expected to be related to Planck
energy. Based on this idea Amelino-Camelia \cite{am} has pioneered
an extended form of SR, popularly known as  Doubly Special theory
of Relativity (DSR). The effect of $\kappa $ appears in the
explicit structures of Lorentz transformations in DSR, which are
non-linear for momenta and momenta-dependent for coordinates. At
the same time $\kappa$ induces a Non-Commutative (NC) spacetime
structure \cite{sn}, which is referred to as the
$\kappa$-Minkowski spacetime \cite{dsr1}. Generally the scale of
$\kappa$ is associated with Planck energy, and all
$\kappa$-induced modifications smoothly disappear in the low
energy sector (or equivalently
in the limit $\kappa \rightarrow \infty $). In the present paper we take $%
\lambda \equiv (1/\kappa )$ as the NC parameter and for $\lambda =0$ one
recovers the commutative limit.

A particular form of DSR extension of (\ref{sr}) that we will be concerned
with in the present paper is given by,
\begin{equation}
p^{2}=M^{2}[1-\lambda E]^{2},  \label{0m}
\end{equation}
with $E$ being the particle energy. This form is known as the
Magueijo-Smolin (MS) construction \cite{mag} of DSR. The corresponding $%
\kappa $-Minkowski NC spacetime \cite{dsr1} is,
\begin{equation}
\{x^{i},x^{0}\}=i\lambda x^{i}~;~~\{x^{i},x^{j}\}=0.  \label{kst}
\end{equation}
We will comment on the classical nature of NC spacetime algebra,
that is being considered here, a little later. In the present
article we will focus on the DSR generalization of Dirac equation
\cite{dsrd} in MS base \cite{mag}. (For alternative constructions
of DSR Dirac equation see \cite{others}.) We will take up directly
the issue of (semi-classical) quantization of the DSR Dirac
particle by exploiting the generalized Foldy-Wouthuysen (FW)
approach \cite{fw}. This extension amounts to a semiclassical
($O(\hbar $) quantization of the Dirac particle in the presence of
external interactions and in previous works of some of us \cite
{ber2} this idea has been successfully employed to reveal
intrinsic Berry phase \cite{berry} effects in the particle
dynamics (see \cite{bl} for a review). The appearance of Berry
phase effects is very natural in FW formalism. The Berry potential
is induced by the electron spin coordinates which are treated as
''fast'' variables as compared to the position coordinates which
are considered as ''slow'' degrees of freedom. The same phenomena
was discovered earlier \cite{mead} when molecular dynamics was
studied in the Born-Oppenheimer approximation. In this context let
us emphasize the novelties in the present work: The DSR Dirac
equation will be derived in an algebraic formalism which is
mathematically very simple and it exploits ideas studied in detail
in a previous work of one of the present authors \cite{sg}.
Interestingly this formalism allows us to introduce interactions
in the DSR scenario in a consistent way. Indeed interactions in a
DSR framework generally have not appeared in the literature.
Lastly the study of DSR Dirac model in FW scheme is entirely new
and (not surprisingly) the results obtained by us can indicate new
directions in this area.

The paper is organized as follows: In Section II we derive the DSR Dirac
equation in Magueijo-Smolin (MS) \cite{mag} base. Section III is devoted to
the FW analysis of the DSR Dirac equation constructed in Section II. In
Section IV we introduce the external interaction and reveal the intrinsic
Berry phase contribution in the present case. The paper ends with
conclusions and areas of future studies in Section V. \vskip .5cm

\begin{center}
DSR DIRAC EQUATION IN MAGUEIJO-SMOLIN BASE
\end{center}

\vskip .2cm Let us first put our approach in its proper
perspective in the context of quantum NC DSR theories. Our aim is
to exploit the canonical framework \cite{sg} in constructing the
theory and subsequently consider its quantization. For this reason
our analysis is completely classical, at least for the time being.
Thus we use (classical) Poisson brackets in place of (quantum)
commutators. On the other hand, the authors of \cite{mpl} discuss
a quantum DSR theory from the very beginning and consider the
spacetime as quantum in nature. However, the classical nature of
the field variables are retained in \cite{mpl} in the sense that
products of fields are not replaced by their $*$-product. In this
section we will construct the DSR Dirac equation in MS base (see
Appendix). To motivate the present derivation, we have to
introduce the full $\kappa$-Minkowski Non-Commutative (NC) phase
space,
\begin{equation}
\{x^i,x^0\}=\lambda x^i ~;~~
\{x^i,x^j\}=0~;~~\{x^i,p^j\}=-g^{ij}~;~\{p^\mu,p^\nu\}=0.  \label{ko}
\end{equation}
\begin{equation}
\{x^0,p^i\}=\lambda p^i ~;~\{x^i,p^0\}=0~;~\{x^0,p^0\}=-1+\lambda p^0.
\label{2}
\end{equation}
We are in the classical framework and will interpret the phase space algebra
as Poisson brackets. Our metric is $diag~g^{00}=-g^{ii}=1$. For convenience
it is expressed in a covariant form,
\[
\{x_\mu ,x_\nu \}=\lambda (x_\mu \eta_{\nu}-x_\nu \eta_{\mu }),
\]
\begin{equation}
\{x_{\mu},p_{\nu}\}=-g_{\mu\nu}+\lambda
\eta_{\mu}p_{\nu},~~\{p_{\mu},p_{\nu}\}=0,  \label{03}
\end{equation}
where $\eta _0=1,\eta _i=0$. For completeness we mention that the Lorentz
generator
\begin{equation}
J_{\mu\nu}=x_{\mu}p_{\nu}-x_{\nu}p_{\mu},  \label{003}
\end{equation}
satisfies the {\it {undeformed}} Lorentz algebra,
\begin{equation}
\{J^{\mu\nu },J^{\alpha\beta }\}=g^{\mu\beta }J^{\nu\alpha }+g^{\mu\alpha
}J^{\beta \nu}+g^{\nu\beta }J^{\alpha\mu }+g^{\nu\alpha }J^{\mu\beta },
\label{51}
\end{equation}
but induces deformations in the transformation laws. The energy-momentum
dispersion law, consistent with the form of $J_{\mu\nu}$ given in (\ref{003}%
) is the MS relation (\ref{0m}),
\begin{equation}
p^2=M^2[1-\lambda E ]^2=M^2[1-\lambda (\eta p) ]^2.  \label{m}
\end{equation}
Indeed in all the above computations one uses the NC phase space algebra (%
\ref{03}).

It is quite obvious that a direct generalization of the physical laws in the
NC phase space is complicated, both mathematically as well as conceptually
\cite{dsrd,others}. However, there is an easy way out \cite{sg} which we now
explain. We can introduce a map $(X_\mu ,P_\mu )=F^{-1}(x_\mu ,p_\mu )$ \cite
{sg} between $(x_{\mu },p_{\mu })$ - the {\it {physical}} $\kappa $%
-Minkowski NC phase space and $(X_{\mu },P_{\mu })$ - a {\it {completely
canonical}} phase space. Explicitly the transformation rules are the
following:
\begin{equation}
X_{\mu }\equiv x_{\mu }(1-\lambda (\eta p))=x_{\mu }(1-\lambda E);~~P_{\mu
}\equiv \frac{p_{\mu }}{(1-\lambda (\eta p))}=\frac{p_{\mu }}{(1-\lambda E)}.
\label{c1}
\end{equation}
The $(X_{\mu },P_{\mu })$ variables obey canonical Poisson Bracket algebra
\begin{equation}
\{X_{\mu },P_{\nu }\}=-g_{\mu \nu };~~\{X_{\mu },X_{\nu }\}=\{P_{\mu
},P_{\nu }\}=0.  \label{c2}
\end{equation}
We have also shown in \cite{sg} that $(X_{\mu },P_{\nu })$ have conventional
(Special Theoretic) Lorentz transformation properties whereas, as mentioned
before, $(x_{\mu },p_{\nu })$ obey $\kappa $-deformed Lorentz transformation
laws \cite{kim}.

The above relations in (\ref{c1}) are invertible,
\begin{equation}
x_\mu = X_\mu (1+\lambda (\eta P))=X_\mu (1+\lambda P_0);~~p_\mu = \frac{%
P_\mu}{(1+\lambda (\eta P))}=\frac{P_\mu}{(1+\lambda P_0)}.  \label{c3}
\end{equation}
We note that the above relations are classical and one has to order them
appropriately under quantization. However, we will show these mappings
survive under some restrictions and in the present work that is sufficient.

Our framework of generalizing physical laws to $\kappa$-Minkowski phase
space is the following: Start with the known form of a relation in the
(auxiliary) canonical phase space $(X_\mu ,P_\nu )$. Now simply map the
relation using (\ref{c1}) to the physical NC $\lambda $-variables $(x_\mu
,p_\nu )$. This yields the cherished $\lambda $-deformed physical law.
Indeed it is not obvious that the procedure will work but we have explicitly
shown its validity in various instances in \cite{sg} in the context of point
particle models. We will demonstrate (see Appendix) that this principle
works for the Dirac equation as well.

To this end, we start with the momentum space Dirac equation in canonical
phase space:
\begin{equation}
(\gamma^{\mu}P_{\mu}-M)u(\vec P)=0,  \label{d1}
\end{equation}
where $\gamma^{\mu}$ are the standard "$\gamma $ matrices". We now provide
the one line derivation of the DSR Dirac equation in MS base:
\begin{equation}
(\gamma^{\mu}\frac{p_\mu}{(1-\frac{(\eta p)}{\kappa})}-M)u(\vec p)=0.
\label{d2}
\end{equation}
This is the $\lambda $-extended Dirac equation in MS basis. This is a new
result. In the Appendix, we substantiate the validity of our claim by
comparing (\ref{d2}) with existing results in the literature \cite{dsrd}.

For applying the FW transformation on our MS-Dirac equation (\ref{d2}) we
express it in Hamiltonian form,
\begin{equation}
\left( E-\vec{\alpha}.\vec{p}-\beta M\left( 1-\lambda E\right) \right) u=0.
\label{d3}
\end{equation}
After rearranging (\ref{d3}) we obtain,
\begin{eqnarray}
E\left( 1+\lambda M\beta \right) u &=&\left( \vec{\alpha}.\vec{p}+\beta
M\right) u \\
Eu &=&\frac{\left( 1-\lambda M\beta \right) }{1-\lambda ^{2}M^{2}}\left(
\vec{\alpha}.\vec{p}+\beta M\right) u.  \label{d4}
\end{eqnarray}
Quite surprisingly we find that the Hamiltonian $H$ (or $E$ in (\ref{d4}))
has a {\it {non-hermitian}} structure:
\begin{equation}
H=A+B\beta +C\vec{\alpha}\vec{p},  \label{d5}
\end{equation}
where
\begin{eqnarray}
A &=&\frac{-\lambda M^{2}}{1-\lambda ^{2}M^{2}}  \label{A} \\
B &=&\frac{M}{1-\lambda ^{2}M^{2}}  \label{B} \\
C &=&\frac{\left( 1-\lambda M\beta \right) }{1-\lambda ^{2}M^{2}}.  \label{C}
\end{eqnarray}
This is a new result. This feature was not revealed in previous analysis
\cite{dsrd,others} since a proper quantum treatment of the DSR Dirac
equation was not attempted. This problem is tackled by performing a
similarity transformation on $H$ with the matrix $D$,
\begin{equation}
D=C^{-1/2}=\sqrt{1+\frac{M}{\kappa }\beta }=a+b\beta ,  \label{d6}
\end{equation}
where
\begin{equation}
a=\sqrt{\frac{1}{2}(1+\sqrt{1-\frac{M^{2}}{\kappa ^{2}}})}~~;~~b=\sqrt{\frac{%
M^{2}}{2\kappa ^{2}(1+\sqrt{1-\frac{M^{2}}{\kappa ^{2}}})}}.  \label{d7}
\end{equation}
This leads to the final Hermitian form of $H$:
\begin{equation}
H=D(A+B\beta +C\vec{\alpha}\vec{p})D^{-1}=A+B\beta +(1-\frac{M^{2}}{\kappa
^{2}})^{-\frac{1}{2}}\vec{\alpha}.\vec{p}.  \label{x}
\end{equation}
This $H$ is suitable for the conventional FW procedure with similarity
transformation performed by the unitary matrix $U$,
\begin{equation}
U=\frac{\sqrt{\left( \frac{1}{1-\lambda ^{2}M^{2}}\right) \vec{p}^{2}+B^{2}}%
+B+\frac{1}{\sqrt{1-\lambda ^{2}M^{2}}}\beta {\bf \alpha }.\vec{p}}{\left( 2%
\sqrt{\frac{\vec{p}^{2}}{1-\lambda ^{2}M^{2}}+B^{2}}\left( \sqrt{\frac{\vec{p%
}^{2}}{1-\lambda ^{2}M^{2}}+B^{2}}+B\right) \right) ^{1/2}},  \label{d10}
\end{equation}
which can also be written as,
\begin{equation}
U=\frac{\left( \sqrt{\left( 1-\lambda ^{2}M^{2}\right) \vec{p}^{2}+M^{2}}+M+%
\sqrt{1-\lambda ^{2}M^{2}}\beta {\bf \alpha }.\vec{p}\right) }{\left( 2\sqrt{%
\left( 1-\lambda ^{2}M^{2}\right) \vec{p}^{2}+M^{2}}\left( \sqrt{\left(
1-\lambda ^{2}M^{2}\right) \vec{p}^{2}+M^{2}}+M\right) \right) ^{1/2}}
\label{e1}
\end{equation}
In this way we see that $U$ is very similar to the usual FW transformation $%
U_{0}$ for a free particle of momentum $\overrightarrow{p}^{2}$ which is
\begin{equation}
U_{0}=\frac{\left( \sqrt{\overrightarrow{p}^{2}+M^{2}}+M+\beta {\bf \alpha }.%
\overrightarrow{p}\right) }{\left( 2\sqrt{\overrightarrow{p}^{2}+M^{2}}%
\left( \sqrt{\overrightarrow{p}^{2}+M^{2}}+M\right) \right) ^{1/2}}.
\label{e2}
\end{equation}
Indeed, $U$ in (\ref{e1}) reduces to $U_{0}$ for $\lambda =0$. {\footnote{%
Once again, in an algebraic way, $U$ can be computed directly from $U_{0}$
by the change of variable $\overrightarrow{p}\rightarrow \sqrt{1-\lambda
^{2}M^{2}}\vec{p}.$ This is simply because in our equation (\ref{x}) the
factor $(1-\lambda ^{2}M^{2})^{-\frac{1}{2}}$ appears in front of $\vec{%
\alpha}.\vec{p}$.}} The energy eigenvalues of $H$ are:
\[
E_{\pm }=A\pm \sqrt{(1-\frac{M^{2}}{\kappa ^{2}})^{-1}\vec{p}^{2}+B^{2}}=%
\frac{-\lambda M^{2}}{1-\lambda ^{2}M^{2}}\pm \sqrt{\left( \frac{1}{%
1-\lambda ^{2}M^{2}}\right) \vec{p}^{2}+\left( \frac{M}{1-\lambda ^{2}M^{2}}%
\right) ^{2}}
\]
\begin{equation}
=\frac{1}{1-\lambda ^{2}M^{2}}\left( -\lambda M^{2}\pm \sqrt{\left(
1-\lambda ^{2}M^{2}\right) \overrightarrow{p}^{2}+M^{2}}\right)  \label{fwd}
\end{equation}

Clearly they satisfy the MS dispersion law (\ref{0m}). This can be
considered as an alternative way of obtaining the MS dispersion law starting
from the MS extension of Dirac equation, in the spirit of \cite{cor}. We
emphasize the framework adopted by us, {\it {i.e.}} constructing the MS
Dirac equation by exploiting the canonical phase space coordinates and
subsequently utilizing the FW formalism to compute the energy eigenvalues
that are identical with the MS energy values, is quantum mechanical in
nature and is totally new.

We would like to make a comment on the particle and anti-particle spectra
corresponding to the MS dispersion law, although this is not directly
relevant for the present work. For the normal Dirac particle in commutative
spacetime, the particle and anti-particle spectra, $E_{\pm }=\pm \sqrt{\vec{p%
}^{2}+M^{2}}$ are symmetrically placed with respect to the zero-energy
level. This is obtained by putting $\lambda =0$ in (\ref{fwd}). At the same
time it is evident that this feature is absent for the MS spectra (\ref{fwd}%
) with a non-zero $\lambda $. However, we can restore this symmetry simply
by {\it {shifting}} the zero level of the energy value by the {\it {constant}%
} amount $\frac{-\lambda M^{2}}{1-\lambda ^{2}M^{2}}$ in (\ref{fwd}). Thus
the particle and anti-particle sector energy levels are given by,
\begin{equation}
E_{\pm }=\pm \frac{\sqrt{(1-\lambda ^{2}M^{2})\vec{p}^{2}+M^{2}}}{1-\lambda
^{2}M^{2}}.  \label{fwd1}
\end{equation}
\vskip.5cm

\begin{center}
PHYSICAL POSITION OPERATOR AND BERRY CURVATURE EFFECTS ON PARTICLE DYNAMICS
\end{center}

\vskip.2cm It is well known that, (in the normal commutative spacetime), the
position operator $\vec{r}$ in Dirac equation does not represent the
physical coordinate of the particle simply because the magnitude of the
particle velocity obtained from taking time evolution of $\vec{r}$ is $c$,
even for a massive particle. On the other hand the FW formalism provides a
natural prescription for a physical position operator $\vec{r}_{+}$ that
correctly reproduces the particle velocity. Incidentally $\vec{r}_{+}$
operators are noncommutative in nature and as we shall see below, this
feature can be attributed to a Berry phase effect. Phenomena of a similar
have been observed before \cite{niu} in condensed matter systems. In the
present case we find that the Berry curvature effect is further modified by
the $\lambda $-corrections.

We now limit ourself to the dynamical evolution of a positive energy
particle whose energy is formally\ the projection (we denote the formal
projection operation by ${\cal P}$) of the Hamiltonian on the positive
energy subspace. Physically, we thus consider the adiabatic approximation
which allows to neglect the interband transition, by identifying the
momentum degree of freedom as slow and the spin degree of freedom as fast,
similarly to the nuclear configuration in adiabatic treatment of molecular
problems. To be coherent with this projection on the positive energy
subspace, the same projection has to be done for all dynamical operators.
Therefore the physical position operator $\overrightarrow{r}_{+}$ is
obtained as
\begin{equation}
\overrightarrow{r}_{+}{\bf =}{\cal P}\left( UD\overrightarrow{r}%
D^{-1}U^{+}\right) ={\cal P}\left( U\overrightarrow{r}U^{+}\right)
\end{equation}
with the MS position operator $\vec{r}=i\hbar \frac{\partial }{\partial \vec{%
p}}$. Notice that we are using the canonical representation of the position
operator, (valid for commutative spacetime), for the $\lambda $-variables as
well. This is allowed if one recalls from (\ref{ko},\ref{2},\ref{03}) that
the spatial sector remains unaffected in this NC spacetime. This can also be
justified in a more rigorous way by going to the canonical phase space
framework (\ref{c1}). This leads to the following explicit form of $\vec{r}%
_{+}$:
\begin{equation}
\overrightarrow{r}_{+}{\bf =}i\hbar {\cal P}\left( U\frac{\partial }{%
\partial \overrightarrow{p}}U^{+}\right) {\bf =}i\hbar \frac{\partial }{%
\partial \overrightarrow{p}}+{\cal P}\left( Ui\hbar \frac{\partial U^{+}}{%
\partial \overrightarrow{p}}\right) \equiv \overrightarrow{r}+{\cal A}_{(r)}
\label{l1}
\end{equation}
where we have introduced ${\cal A}_{(r)}$, the so called Berry connection.
The $(r)$ in ${\cal A}_{(r)}$ indicates that this Berry connection
accompanies the position $\overrightarrow{r}$. (In general, Berry
connections can appear both in coordinate and momentum.) Indeed the state
space of the Dirac electron is spanned by the basis of plane waves of the
form $\left| p,\varphi _{\alpha }\right\rangle =\left| p\right\rangle \left|
\varphi _{\alpha }\right\rangle $, where $\left| \varphi _{\alpha
}\right\rangle $ is a zero-momentum spinor, such that $U(p)\left| \varphi
_{\alpha }\right\rangle =\left| u_{\alpha }(p)\right\rangle $ which is a
spinor solution of the Dirac equation. For $\left| \varphi _{\alpha
}\right\rangle =(1,0,0,0)$ and $\left| \varphi _{\beta }\right\rangle
=(1,0,0,0)$ the matrix elements of ${\cal A}_{(r)}$ are then
\[
\left( {\cal A}_{(r)}\right) _{\alpha \beta }=i\hbar \left\langle \varphi
_{\alpha }\right| U\left( p\right) \frac{\partial U^{+}\left( p\right) }{%
\partial \overrightarrow{p}}\left| \varphi _{\beta }\right\rangle =i\hbar
\left\langle u_{\alpha }(p)\right| \frac{\partial }{\partial \overrightarrow{%
p}}\left| u_{\beta }(p)\right\rangle .
\]
This is the definition of the Berry connection. The explicit structure of $%
{\cal A}_{(r)}$ in the present context is,
\begin{equation}
{\cal A}_{(r)}={\cal P}\left( U\overrightarrow{r}U^{-1}\right) =\hbar \frac{%
\sqrt{\left( 1-\lambda ^{2}M^{2}\right) }(\vec{p}\times \overrightarrow{{\bf %
\sigma }})}{2\sqrt{\left( 1-\lambda ^{2}M^{2}\right) \vec{p}^{2}+M^{2}}%
\left( \sqrt{\left( 1-\lambda ^{2}M^{2}\right) \vec{p}^{2}+M^{2}}+M\right) },
\label{b2}
\end{equation}
with $\overrightarrow{{\bf \sigma }}$ being the Pauli matrices. Note that
for a free particle the momentum is FW invariant, i.e. $\overrightarrow{p}%
_{+}{\bf =}{\cal P}\left( U\overrightarrow{p}U^{-1}\right) =\overrightarrow{p%
}.$

As we have mentioned before, the physical position operators $\vec{r}_{+}$
are no longer commutative:
\begin{eqnarray}
\left[ r_{+}^{i},r_{+}^{j}\right] &=&i\hbar \varepsilon ^{ijk}\Theta
_{k}=i\hbar \left( \left( \nabla ^i _{(p)}{\cal A}^j_{(r)}-\nabla ^j _{(p)}%
{\cal A}^i_{(r)}\right) +\left[ {\cal A}^i_{(r)},{\cal A}^i_{(r)}\right]
\right)  \nonumber \\
\left[ p^{i},p^{j}\right] &=&0  \nonumber \\
\left[ p^{i},r_{+}^{j}\right] &=&-i\hbar \delta ^{ij}
\end{eqnarray}
with the {\it {non-abelian}} Berry curvature defined by
\begin{equation}
\overrightarrow{\Theta }=-\frac{\hbar }{\left( \left( 1-\lambda
^{2}M^{2}\right) \vec{p}^{2}+M^{2}\right) ^{3/2}}\left( M\overrightarrow{%
\sigma }+\frac{\left( 1-\lambda ^{2}M^{2}\right) \left( \vec{p}.%
\overrightarrow{\sigma }\right) \vec{p}}{\sqrt{\left( 1-\lambda
^{2}M^{2}\right) \vec{p}^{2}+M^{2}}+M}\right) .  \label{b4}
\end{equation}
The dynamics is generated by the Hamiltonian equations of motion,
\begin{equation}
\dot{O}=i[H,O],  \label{b5}
\end{equation}
where $O$ is a generic operator and $H$ is the Hamiltonian. In the present
instance with $E_{+}$ representing $H$, we find,
\begin{equation}
\stackrel{\cdot }{\overrightarrow{r}}_{+}=\frac{\overrightarrow{p}}{\sqrt{%
\left( 1-\lambda ^{2}M^{2}\right) \overrightarrow{p}^{2}+M^{2}}}~~;~~%
\stackrel{\cdot }{\overrightarrow{p}}=0.  \label{b6}
\end{equation}
Also one finds that the non-relativistic expression for velocity is obtained
for $M=1/\lambda $ which in fact corresponds to the upper bound of mass (or
energy) available for the particle. In the next section we will study the
dynamics of the MS Dirac particle in presence of external interaction. \vskip%
.5cm.

\begin{center}
DSR DYNAMICS IN PRESENCE OF SCALAR INTERACTION AND BERRY CURVATURE EFFECTS
\end{center}

\vskip.2cm

In general it is not possible to perform the FW transformation exactly when
interactions are present but one can always resort to a perturbative study
for a ''small'' potential term. But here instead of a perturbative expansion
we consider a semiclassical approximation ($O(\hbar )$) and in addition
limit ourself to $O(\lambda )$ effect only.

Once again we start with the free MS Dirac equation (\ref{d2}) in canonical
phase space and introduce an interaction potential $V(\vec{R})$,
\begin{equation}
\{\gamma ^{\mu }P_{\mu }-M-\gamma ^{0}V(\vec{R})\}u(\vec{P})=0.
\end{equation}
This interaction can be thought of as scalar component of $U(1)$
interaction, in a particular frame where the vector potential vanishes.

Translated to $\lambda $-variables this equation reads,
\begin{equation}
\{\gamma ^{\mu }\frac{p_{\mu }}{(1-\lambda (\eta p))}-M-\gamma ^{0}V(\vec{r}%
(1-\lambda E))\}u(\vec{p})=0.
\end{equation}
Let us consider the $\lambda $ in its lowest non-trivial order. Hence to $%
O(\lambda )$ the Dirac equation is,
\begin{equation}
\left( 1+\lambda {V\left( \overrightarrow{r}\right) +}\lambda {%
\overrightarrow{r}.{\bf \nabla }V\left( \overrightarrow{r}\right) }\right)
E~u(\vec{p})=(A+B^{\prime }\beta +C(\vec{\alpha}.\vec{p})+V\left(
\overrightarrow{r}\right) )u(\vec{p}).  \label{b7}
\end{equation}
with $B^{\prime }=B-\lambda MV\left( \overrightarrow{r}\right) $ and $A,$ $B$%
, $C$ given in Eqs.(\ref{A}-\ref{C})

To fit in the FW scheme (\ref{b7}) is rewritten as,
\begin{equation}
Eu(\vec{p})=\left( 1-\lambda {V\left( \overrightarrow{r}\right) }-\lambda {%
\overrightarrow{r}.{\bf \nabla }V\left( \overrightarrow{r}\right) }\right)
(A+B^{\prime }\beta +C(\vec{\alpha}.\vec{p})+V\left( \overrightarrow{r}%
\right) )u(\vec{p}).  \label{b8}
\end{equation}
The same procedure as in the free case is needed to render this $H$
hermitian and we obtain a Hamiltonian that is hermitian in $O(\lambda )$ and
$\hbar $:
\begin{eqnarray}
H &=&D(1-\lambda {V\left( \overrightarrow{r}\right) -}\lambda
\overrightarrow{r}.{\bf \nabla }V\left( \overrightarrow{r}\right)
)(A+B^{\prime }\beta +C\vec{\alpha}.\vec{p}+V\left( \overrightarrow{r}%
\right) )D^{-1}  \nonumber \\
&=&\left( 1-\lambda {V\left( \overrightarrow{r}\right) }-\lambda
\overrightarrow{r}.{\nabla }V\left( \overrightarrow{r}\right) \right)
D(A+B^{\prime }\beta +C\vec{\alpha}\vec{p}+V\left( \overrightarrow{r}\right)
)D^{-1}  \nonumber \\
&=&\left( 1-\lambda {V\left( \overrightarrow{r}\right) }-\lambda
\overrightarrow{r}.{\nabla }V\left( \overrightarrow{r}\right) \right) \left(
A+B^{\prime }\beta +(1-\frac{M^{2}}{\kappa ^{2}})^{-\frac{1}{2}}\vec{\alpha}.%
\vec{p}.+V\left( \overrightarrow{r}\right) \right)
\end{eqnarray}
In ref \cite{Method}{\bf \ }we have shown how to diagonalize Dirac like
Hamiltonians in the presence of external fields at the semiclassical order (
to get corrections beyond the semiclassical order see ref \cite{allorder}).
During this process of diagonalization, we have shown that in general (in
the presence of electromagnetic or gravitational fields), both position and
momentum operators acquire a Berry-phase contribution making the coordinate
and momentum algebra noncommutative. Note that in the presence of
interactions, the projected (on the positive energy subspace) dynamical
operators emerge naturally during the diagonalization, without resorting to
the adiabatic approximation.

Therefore applying this diagonalization procedure in our case leads for the
positive energy sector to the following semiclassical diagonal Hamiltonian \

\begin{equation}
H_{+}={\cal P}\left( UHU^{-1}\right)
\end{equation}
where $U$\ is the same matrix as in the free case with $M$\ replaced by an
effective mass{\bf \ }$M\left( 1-\lambda V\left( \overrightarrow{r}\right)
\right) $. This leads to
\begin{eqnarray}
H_{+} &=&\left( 1-\lambda {V\left( {\overrightarrow{r}}_{+}\right) -}\lambda
{\overrightarrow{r}}_{+}{{\bf \nabla }V\left( \overrightarrow{r}_{+}\right) }%
\right) \left( \sqrt{\overrightarrow{p}^{2}+\left( 1-\lambda V\left(
\overrightarrow{r}_{+}\right) \right) ^{2}M^{2}}-\lambda M^{2}+V\left(
\overrightarrow{r}_{+}\right) \right)   \nonumber \\
&&+o(\sqrt{\lambda ^{2}+\hbar ^{2}})
\end{eqnarray}
where $\overrightarrow{r}_{+}=\overrightarrow{r}+{\cal A}_{(\lambda ,r)}$
with ${\cal A}_{(\lambda ,r)}$ the same as in the free case with the
effective mass $M\left( 1-\lambda V\left( \overrightarrow{r}_{+}\right)
\right) .$ We observe a renormalization of the energy through the term
\[
\left( 1-\lambda {V\left( {\overrightarrow{r}}_{+}\right) -}\lambda {%
\overrightarrow{r}}_{+}{.{\bf \nabla }V\left( \overrightarrow{r}_{+}\right) }%
\right) .
\]
Note that the momentum get also a Berry potential correction which at the
order considered is negligeable $\overrightarrow{p}_{+}=\overrightarrow{p}%
+O(\lambda \hbar ).$ The commutation relations are also unchanged exept for
the mass renormalization. Hence, the dynamics in presence of the potential
is derived to be,
\begin{equation}
\stackrel{\cdot }{\overrightarrow{r}}_{+}=\nabla _{(p)}H_{+}-\stackrel{\cdot
}{\overrightarrow{p}}{\bf \times \Theta }~~;~~\stackrel{\cdot }{%
\overrightarrow{p}}=-\nabla _{(r_{+})}H_{+},  \label{b11}
\end{equation}
where ${\bf \Theta }$ is given in (\ref{b4}) (with $M$\ replaced by an
effective mass{\bf \ }$M\left( 1-\lambda V\left( \overrightarrow{r}\right)
\right) $ ). Note that the effect of Berry curvature is manifested very
directly once an interaction is present. Indeed, the equation for the
velocity contains an anomalous velocity term $\stackrel{\cdot }{%
\overrightarrow{p}}{\bf \times \Theta }$ of order $\hbar $ which causes an
additional displacement of the electrons orthogonally to the momentum $%
\overrightarrow{p}$. This phenomenon depends on the particle spin through
the Berry curvature. Typically, in canonical spacetime, with $V$ being an
electrostatic potential, this is the well studied spin Hall effect. In fact
it should be mentioned that the topics of spin Hall effect of electrons or
photons, gravitational Hall effect for photons, that have become areas of
intense research, all owe there existence to this type of Berry curvature
effect. However, as we point out below, modelling the analogue of spin Hall
effect for MS particle is probably more complicated. \newline
\vskip.5cm

\begin{center}
CONCLUDING REMARKS
\end{center}

\vskip.2cm In this paper we have considered particle dynamics in $\kappa $%
-Minkowski spacetime which is a particular form of noncommutative
spacetime. Indeed quantization of the particle model is tricky
because of the operatorial form of noncommutativity involved in
the phase space commutation relations. The novelty of our work
lies in the fact that we have been able to bypass this problem by
exploiting a semiclassical approach of the Foldy-Wouthuysen
formalism. Indeed this makes our analysis semi-classical in nature
with quantum effects in $(O(\hbar )$\ taken into account. We then
have shown that the dynamics of MS particles in the presence of an
external field should be influenced by a the $\kappa $-parameter
but also by the presence of a Berry curvature which is itself
$\kappa $ dependent.

In this connection, let us pause to mention the significant (and
possibly more ambitious) work \cite{int} that attempts to deal
with the full  $\kappa$-Minkowski quantum field theory.
Considering the $*$-product for the fields, the authors of
\cite{int} demonstrate that if properly interpreted, the
interaction vertices have the requisite symmetry under the
interchange of momenta of identical incoming particles. Indeed,
the $\kappa$-Minkowski $*$-product is much more involved that
(Moyal) $*$-product that appears when the non-commutativity is
constant in nature. It is true that we have so far only looked at
the quantum mechanics of a DSR model but the nature of the
energy-momentum conservation laws \cite{sg} inherently possess the
above mentioned symmetry. This seems to indicate that the DSR in
Magueijo-Smolin base might be a better alternative that DSR in
other bases in the context of formulating the DSR quantum field
theory.

The next task we wish to pursue is the effect of full electromagnetic
interactions on the DSR particle model considered here. This is a
non-trivial extension since electromagnetism involves a $U(1)$ gauge
symmetry and one has to appropriately generalize the concept of gauge
invariance in noncommutative spacetime. This feature probably did not show
up in the present restricted setup.

\vskip .5cm

\begin{center}
APPENDIX:
\end{center}

\vskip .2cm We will compare the MS Dirac equation obtained here in (\ref{d2}%
) with that derived in \cite{dsrd}. It is to be noted that \cite{dsrd} uses
the bicrossproduct base whereas we have used the MS base.

It is well-known in the DSR community that there are distinct formulations
(or bases) of the DSR regarding the modified dispersion relation,
transformation properties and phase space algebra. The most popular are
known as bicrossproduct base \cite{dsr1}, MS base \cite{mag} and standard
base \cite{dsr2}. The different bases are connected by non-linear
transformation and \cite{kow} but the different bases are inequivalent with
drastically different physical consequences due to the non-linearity in the
transformations.

The DSR Dirac equation obtained in \cite{dsrd} is
\begin{equation}
(\gamma ^{\mu }D_{\mu }-\omega )u=0,  \label{a}
\end{equation}
where,
\begin{eqnarray}
D_{0} &=&\frac{e^{\lambda E^{\prime }}-\cosh (\lambda \omega )}{\sinh (\frac{%
\omega }{\kappa })} \\
D_{i} &=&\frac{e^{\frac{E^{\prime }}{\kappa }}}{\kappa \sinh (\lambda \omega
)}p_{i}^{\prime }.
\end{eqnarray}
Here $(E^{\prime },p_{i}^{\prime })$ are energy and momenta in
bicrossproduct basis and they satisfy the dispersion law \cite{dsr1},
\begin{equation}
\frac{2}{\lambda ^{2}}cosh(\lambda E^{\prime })-(\vec{p}^{\prime
})^{2}e^{\lambda E^{\prime }}=2\kappa ^{2}cosh(\lambda \omega ).  \label{a1}
\end{equation}
The relations connecting the above set to the MS set of variables (that we
have used) are \cite{kow},
\begin{equation}
p_{i}^{\prime }=p_{i}~;~E^{\prime }=-\frac{1}{2\lambda }log(1-2\lambda
E+\lambda ^{2}\vec{p}^{2}),E=\frac{1}{2\lambda }(1-e^{-2\lambda E^{\prime
}}+\lambda ^{2}\vec{p}^{\prime 2})~;~\lambda m=tanh(\lambda \omega ).
\label{a2}
\end{equation}
It is straightforward to check that using (\ref{a2}) on (\ref{a})
will generate (\ref{d2}), the MS Dirac equation derived in our
paper. \vskip.5cm {\it {Acknowledgement}}: S.G. would like to
thank the Physics Department, Paul Verlaine-Metz University,
France, for their very kind and congenial hospitality. We are
grateful to the Referee for the constructive and detailed
observations.

\end{document}